\documentclass[11pt]{article}

\begin{document}
\begin{center} {\large \bf Spectral Modelling of Quantum
Superlattice and Application to the Mott-Peierls Simulated Transitions }
\end{center}

 \begin{center} L.A.Dmitrieva$^{1)}$, Yu.A.Kuperin$^{2)}$
\end{center}
\vskip0.2cm
{\small \begin{center}
1) Division of Mathematical and Computational Physics,\\
St.Petersburg State University,
198504 St.Petersburg, Russia\\
mila@JK1454.spb.edu

\vskip0.2cm
2) Laboratory of Complex Systems Theory,\\
St.Petersburg State University,
St.Petersburg 198504, Russia\\
kuperin@JK1454.spb.edu
\end{center}}

\vskip0.3cm
\begin{abstract}
A local perturbation theory for the spectral analysis of the
Schr\"odinger operator with two periodic potentials
whose periods are commensurable has been constructed. It has
been shown that the perturbation of the periodic 1D Hamiltonian
by an additional small periodic potential leads to the following
spectral deformation: all gaps in the spectrum of the unperturbed
periodic Hamiltonian bear shifts while any band splits by
arising additional gaps into a set of smaller spectral bands.
The spectral shift, the position of additional gaps and their
widths have been calculated explicitly. The applications to the
operational regime of a nanoelectronic device based on Mott-Peierls
stimulated transition have also been discussed.

\end{abstract}

\section{Introduction}

In series of papers \cite{1,2,3,Anton,4,5} a new type of nanoelectronic device
based on the phenomenon of the stimulated Mott-Peierls transition
(SMPT) has been studied in one electron and one-mode
approximation. The spectral properties and the operational
regimes of the device have been calculated for the ballistic
electrons moving along the quantum wire (QW) treated as
the conducting electrode of the device. The effect of
QW-structure has been taken into account by the effective mass
$m^*$ in the Schr\"odinger equation
\begin{equation} \label{I.1}
-\frac{\hbar^2}{2m^*} \frac{d^2 \Psi}{dx^2}\, = \,E\Psi
\end{equation}
under assumption that normally the conducting electrode is in
metallic state. When a nontrivial operating voltage has been
applied to the governing electrodes distributed periodically
along the QW it has been shown that the effect of the governing
electrodes can been reduced to an effective periodic potential
$V_{eff}$ in the Schr\"odinger operator (\ref{I.1}):
\begin{equation} \label{I.2}
-\frac{\hbar^2}{2m^*} \frac{d^2 \Psi}{dx^2}\, + \,V_{eff} \Psi
\,\equiv \,H_{eff} \,= \,E\Psi.
\end{equation}
The spectrum of the operator (\ref {I.2}) consists of spectral bands
separated by gaps. The device is in conducting state if the
Fermi level of the conducting electrode gets into one of the
artificially created spectral bands. On the contrary, if the
Fermi level gets into a gap between the spectral bands the
conductivity vanishes and thus the conducting electrode is
effectively transformed into the dielectric state. Summarizing
what has been said one can conclude that the formation of the
energy gaps in the continuous spectrum of the operator (\ref {I.1}) and
the control of the parameters of $V_{eff}$ may be considered as
the physical base of a new type of nanoelectronic device.

For a more detailed analysis of spectral properties and physical
characteristics (say, conductance) of the device described above
deviations from the {\it ideal} behavior (one-electron,
one-mode ballistic regime at zero temperature) should be
considered because the effect of electron-phonon interaction,
electron-electron interaction, stimulated superlattice structure of
governing electrode , etc., may become important for real
device. In particular, if the electron concentration is large
enough, electrons may be effected by a atomic lattice structure
of a QW that could change the behavior of the conductance. Also
if an artificial lattice (periodic) structure could be
fabricated for quantum wire then two periodic structures (atomic
and artificial) could change the spectral properties of the
device as well as its operational regimes.

In the present paper we study the spectral properties of the
controlled QW which is described by 1D quantum Hamiltonian
\begin{equation} \label{I.3}
H\,=\,-\frac{d^2}{dx^2}\,+\,q_0(x)\,+\,\delta
q_1(x)
\end{equation}
with the periodic potentials $q_0(x)$ and $q_1(x)$ playing the
role of the atomic potential of QW in one-electron approximation
$(q_0)$ and the effective potential $(q_1)$ generated by the
periodic system of governing electrodes. Everywhere below we set
$\hbar/2m = 1$ where $m$ is the electron mass and in these units the
parameter $\delta$ is proportional to the ratio of $q_0$ and
$q_1$ amplitudes. We assume that the periods $a$ and $b$ of $q_0$
and $q_1$ respectively are commensurable and we shall call such
QW as 1D superlattice. We consider $\delta$ as a small parameter
which is in agreement with strength of the atomic potential in
comparison with the amplitude of the $V_{eff}$ calculated in
\cite {3}. To study the spectrum of the Hamiltonian (\ref{I.3}) and its
eigenfunctions we develop a special local perturbation theory
which allows to calculate all spectral ingredients of $H$ not
only for regular points but also for singular ones. The latter
as it is shown in the paper coincide with the ends of spectrum
of the operator $H_0$:
\begin{equation} \label{I.4}
H_0\,=\,-\frac{d^2}{dx^2}\,+\,q_0(x)
\end{equation}
and with some set of points inside the bands of $H_0$. This set
is described in the frame of our approach explicitly. We prove
that the main spectral effect of the perturbation is the
following deformation of $\sigma(H_0)$: all gaps of $H_0$ bear
shifts while any spectral band of $H_0$ splits by the arising
additional gaps into a number of smaller spectral bands of $H$.
It is shown that spectral shifts, the locations of additional
gaps and their widths can be calculated in arbitrary order of
the perturbation theory explicitly. In this respect the spectral
analysis presented here can be considered as the sound mathematical
background for numerical calculations of real nanoelectronic
devices fabricated on the basis of narrow-gap semiconductors \cite{2}.
The calculations of the spectral characteristics
and operational regimes for real SMPT-device will be done elsewhere.

\section{The Perturbation Theory at Regular Points}
We consider the perturbation of the Schr\"odinger operator
with a periodic potential by another periodic potential
with a small amplitude. So the Hamiltonian has the form
$$
H\,=\,-\partial_x^2\,+\,q_0(x)\,+\,\delta
q_1(x),\,\,\,x\,\in\,{\bf R},
$$
where $q_0(x+a)=q_0(x)$ and $q_1(x+b)=q_1(x)$ are two
periodic functions, the parameter $\delta$ is considered to
be small. We assume that the periods $a$ and $b$ are
commensurable, i.e.
$ b/a =  m/n$ where $m$ and $n$ are integers. Hence, the perturbed
Hamiltonian $H$ remains to be the operator with a periodic
potential and
$c=bn=am$
is the common period of functions $q_0$ and $q_1$.

In what follows we need some well known facts \cite{6,7} from the
theory of the periodic Schr\"odinger operators $H_{per}$. The spectrum
of such operators consists of spectral bands separated by
gaps. In the general case there are infinite number of bands
and gaps. If the period of the potential is A then the so-called
Bloch functions of $H_{per}$ are solutions of equation
$H_{per}\Psi = E\Psi$ which obey the quasiperiodic boundary
conditions on the period
$\Psi_{\pm}(x+A,k)=e^{\pm ikA}\Psi_{\pm}(x,k).$
The quasimomentum $k$ is linked with the energy $E$ by the
dispersion function $E=E(k)$. Under this mapping to the $N-th$ band
of $H_{per}$ there corresponds the Brilloin zone $[\pi (N-1)/A,
\pi N/A]$ of the values of quasimomentum. To the N-th gap there
corresponds the cut $[\pi N/A, \pi N/A + i\gamma_N]$ drawn on the
complex plane of quasimomentum. The so-called Lyapounov function
$F(E) = \cos kA$ is the entire function of $E$, $|F(E)| < 1$
in bands and $|F(E)| > 1$ in gaps. At the end of bands
$\frac{d}{dk} E(k) = 0$ and at the gap points $E^{(N)}$ corresponding
to the tops $\pi N/A + i\gamma_N$ of cuts the relation
$\frac{d}{dE} F(E) = 0$ is valid.

We assume that the location of bands of the unperturbed
operator $H_0 = -\partial_x^2 + q_0(x)$, its Bloch functions and
the dispersion relation are known. Our aim is to study the
corresponding quantities for the perturbed operator $H$ and to describe
its spectrum in comparison with the spectrum of the unperturbed operator.

Let us present the Bloch solutions $\Psi(x,p)$ of the
perturbed spectral problem
\begin{equation} \label{1}
H\Psi = E\Psi
\end{equation}
in the form
\begin{equation} \label{2}
\Psi(x,p) = e^{i(p-k)x}\Phi(x,p),
\end{equation}
where $E=E(p)$ and $p=p(E)$ are the dispersion function and
the quasimomentum corresponding to the operator $H$,
$\lambda=\lambda(k)$ and $k=k(\lambda)$ are the analogous
objects for the unperturbed problem
\begin{equation} \label{3}
H_0\Psi_0 = \lambda \Psi_0,\,\,\,\,H_0 = -\partial_x^2 +
q_0(x).
\end{equation}

Due to Floquet theorem the function $e^{-ikx}\Phi(x,p)$ is
the periodic one with a period $c$. Hence the function
$\Phi(x,p)$ is quasiperiodic and
\begin{equation} \label{4}
\Phi(x+c,p) = e^{ikc}\Phi(x,p).
\end{equation}

To fix the Bloch solution $\Psi(x,p)$ and hence the
function $\Phi(x,p)$ unambiguously we normalize it as
follows
\begin{equation} \label{5}
\Psi(0,p) = 1, \,\,\,\Psi(c,p) = e^{ipc}.
\end{equation}
Obviously,
$$
\Phi(0,p) = 1,\,\,\,\Phi(c,p) = e^{ikc}.
$$

On use of eqs.(\ref{1}) and (\ref{2}) one can see that the function
$\Phi(x,p)$ satisfies the equation
\begin{equation} \label{6}
\left[{-\partial_x^2 - 2i(p-k)\partial_x + (p-k)^2 + q_0(x)
+ \delta q_1(x) -E}\right]\Phi(x,p) = 0.
\end{equation}
This equation is to be solved in the class of functions
with the boundary conditions (\ref{4}).

Let us seek the solution of eq. (\ref{6}) in the form

\begin{equation} \label{7}
\Phi(x,p) = \Psi_0(x,k) + \mathop{\sum}\limits_{n\geq
1}\delta^n\Psi_n (x,k)
\end{equation}
assuming that the perturbed quasimomentum $p$ and the
dispersion function $E=E(p)$ are also expanded in series
\begin{equation} \label{8}
p = k + \mathop{\sum}\limits_{n\geq
0}\delta^n\mu_n (k)
\end{equation}
and
\begin{equation} \label{9}
E(p) = \lambda(k) + \mathop{\sum}\limits_{n\geq
0}\delta^n\lambda_n (k),
\end{equation}
where $k$ and $\lambda(k)$ are the quasimomentum and
the dispersion function corresponding to the unperturbed
problem (\ref {3}).

Since the unperturbed Bloch function $\Psi_0(x,k)$
satisfies the condition
$
\Psi_0(x+a,k)=e^{ika}\Psi_0(x,k)
$
and hence
$
\Psi_0(x+c,k)=e^{ikc}\Psi_0(x,k),
$
in order to provide the boundary condition (\ref{4}) one has to
seek the correction terms under the same conditions, i.e.
\begin{equation} \label{10}
\Psi_n(x+c,k)=e^{ikc}\Psi_n(x,k),\,\,\,n \geq 1.
\end{equation}

The normalization conditions
\begin{equation} \label{11}
\Psi_0(0,k) = 1,\,\,\,\,\Psi_0(a,k) = e^{ika},
\end{equation}
\begin{equation} \label{12}
\Psi_n(0,k) = \Psi_n(c,k) = 0,\,\,\,n \geq 1
\end{equation}
provide the property (\ref {5}) of the exact Bloch solution.

In what follows we propose a procedure of
calculation of all correction terms $\Psi_n(x,k)$,
$\mu_n(k)$, $\lambda_n(k)$ and define the values of
unperturbed quasimomentum $k$ at which this procedure
fails. The corresponding energy points $\lambda_s =
\lambda(k_s)$ we shall call as singular points of perturbation
series constructed here. The complement of a set of singular
points on real axis we shall call as a set of regular points.
In the next Section we construct the local
perturbation series which are valid at the above singular
points.

The procedure of calculating the correction terms
$\Psi_n(x,k)$, $\mu_n(k)$ and $\lambda_n(k)$ includes
the following steps:

1. For each $n\geq 1$ we find the linear link between the
terms $\lambda_n(k)$ and $\mu_n(k)$ in expansions (\ref{8}) and
(\ref{9}). To this end we use the fact that these expansions are
not independent and are linked with each other by means of
equation
\begin{equation} \label{13}
{\cal F}(E) = \cos pc
\end{equation}
where ${\cal F}(E)$ is the Lyapounov function of the
perturbed operator $H$.

2. For each $n\geq 1$ we find the second linear link
between $\lambda_n(k)$ and $\mu_n(k)$ which is necessary to
define $\lambda_n(k)$ and $\mu_n(k)$ unambiguously. This
second link is obtained by examining the solvability
condition of the inhomogeneous equation for
$\Psi_n(x,k)$
in the class of functions with boundary conditions (\ref{10}).

3. Finally we find the solution $\Psi_n(x,k)$ which satisfies
the boundary conditions (\ref{10}) and (\ref{12}).

All technical details of the realization of the above three
steps are contained in the proofs of statements formulated
below.

\underline{\bf Proposition 1}. {\it The Lyapounov function
${\cal F}(E)$ associated with the
operator $H$ in the leading order of $\delta$ can be expressed
in terms of the unperturbed quasimomentum $k$ as follows}
$
{\cal F}(E) \,=\, \cos kc \,+\,O(\delta).
$

\underline{\it Proof}. Let us insert the expansion (\ref{9}) into the
r.h.s. of the dispersion relation (\ref{13}) and expand the
r.h.s. in the Taylor series at the vicinity of k.
If we restrict the Taylor series to the leading order we obtain
the statement of proposition. Q.E.D.

\underline{\it Remark}. It should be noted that the leading term $\cos
kc$ of the Lyapounov function ${\cal F}(E)$ does not
coincide with the Lyapounov function $F_0(\lambda) = \cos
ka$ of the unperturbed operator $H_0$. This fact has crucial
consequences for the spectrum of the perturbed operator
which is to be discussed below.

To formulate the next statement we need the following
notations:
$
F(\lambda)\vert_{\lambda=\lambda(k)} \equiv \cos kc,
$
$
F' =
\frac{d}{d\lambda}F(\lambda),\,\,\,
F^{(n)}=\frac{d^n}{d\lambda^n}F(\lambda),\,\,\,
\dot{\lambda}(k) = \frac {d}{dk}\lambda (k).
$

\underline{\bf Proposition 2}. {\it The correction terms of
series (\ref{9}) and (\ref{8}) for $E(p)$ and $p$ are linked by the
recursion relations}
\begin{equation} \label{14}
\dot{\lambda}(k)\mu_n(k) - \lambda_n(k) =
T_n\{\lambda_i,\mu_i ,i\leq n-1\}
\end{equation}
{\it where $T_1 = const$ and $T_n\{\cdot,\cdot\}, n>1,$
are some functions of $\lambda_i,\,\mu_i, i\leq n-1,$ which can
be calculated explicitly for arbitrary $n$. In particular, the
first $T_n, n=1,2,3,$have the form}
$$
T_1\,=\, 0,
$$
$$
T_2(\lambda_1, \mu_1)\, =\, \frac12
\left({\frac{F''}{F'} \lambda_1^2 -
\frac{F}{F'}c^2\mu_1^2}\right),
$$
$$
T_2(\lambda_1, \mu_1, \lambda_2, \mu_2)\,=
\,\frac{\dot{\lambda}(k)c^2\mu_1^3}{6} +
\frac{F''}{F'}\lambda_1\lambda_2
+ \frac{F'''}{6F'}\lambda_1^3 + \frac{F}{F'}\mu_1\mu_2c^2.
$$

\underline{\it Proof}. To obtain the above relations one has
to insert expansions (\ref{9}) and (\ref{8}) for $E$ and $p$ into the
dispersion relation (\ref{13}), to expand the l.h.s into the Taylor
series at the vicinity of $\lambda$, to expand the r.h.s. into
the Taylor series at the vicinity of $k$ and to equate coefficients at
each power of $\delta$. Finally one has to take into account that in the
leading order ${\cal F}(E) = F(\lambda)$. Q.E.D.

The $n$-th relation in (\ref{14}) gives one linear link between two
quantities $\lambda_n(k)$ and $\mu_n(k)$. To define them
unambiguously one needs the second link between them. To
obtain it let us consider the solvability condition of the
equation defining the $n$-th correction term $\Psi_n(x,k)$
of the Bloch function. To write down this equation let us
insert the expansions (\ref{7}), (\ref{8}) and (\ref{9})
into the exact equation
(\ref{6}) and equate the terms at each power of $\delta$. As the
result one obtains the set of equations
\begin{equation} \label{15}
\left[{-\partial_x^2 + q_0(x) - \lambda(k)}\right]
\Psi_n(x,k) = R_n(x,k),
\end{equation}
where the inhomogeneous term $R_n(x,k)$ can be expressed in terms
of functions $\Psi_i(x,k)$ at $i \leq n-1$, $\lambda_i(k)$
and $\mu_i$ at $i \leq n$ and the potential $q_1(x)$. For
instance, the  first two terms have the form
$$
R_1(x,k) = \left[{\lambda_1(k) - q_1(x) +
2i\mu_1(k)\partial_x}\right] \Psi_0(x,k),
$$
$$
R_2(x,k) = \left[{\lambda_1(k) - q_1(x) +
2i\mu_1(k)\partial_x}\right] \Psi_1(x,k) +
$$
$$
+ \left[{\lambda_2(k) - \mu_1^2(k) +
2i\mu_2(k)\partial_x}\right] \Psi_0(x,k)
$$
Equation (\ref{15}) is to be solved in the class of functions with
quasiperiodic boundary conditions
\begin{equation} \label{16}
\Psi_n(x+c, k) = e^{ikc}\Psi_n(x,k).
\end{equation}

According to the Fredholm theorem the solvability conditions of
eq. (\ref{15}) is the orthogonality of the r.h.s. of (\ref{15}) to the
solutions of the corresponding homogeneous equation
\begin{equation} \label{17}
\left[{-\partial_x^2 + q_0(x) - \lambda(k)}\right]
\Psi_n^{(0)}(x,k) = 0
\end{equation}
with the same boundary conditions (\ref{16}). It is obvious that the
unique solution of eq. (\ref{15}) satisfying the boundary
conditions (\ref{16}) is given by the Bloch function $\Psi_0(x,k)$.
Thus the solvability conditions have the form
\begin{equation} \label{18}
\langle R_n,\Psi_0 \rangle \equiv \mathop{\int}_0^c R_n(x,k)
\bar\Psi_0(x,k) dx = 0.
\end{equation}
The n-th solvability condition from (\ref{18}) gives the second linear link
between $\lambda_n(k)$ and $\mu_n(k)$.

Before writing down it explicitly let us note that the
solvability conditions of equations similar to (\ref{15}) has been
used in \cite {B1, B2} for constructing the quasiclassical series
in the problem of the adiabatic perturbation of the periodic
potential. In the present paper we construct the perturbational
series, however the specific character of perturbation leads to
the necessity of using the technical tools which are more
appropriate for the quasiclassical approach.

Turn back to the link between $\lambda_n(k)$ and $\mu_n(k)$
and introduce the following functions of k:
$$
N_n(k) = \mathop{\int}_0^c\Psi_n(x,k)\bar\Psi_0(x,k) dx,
\,\,\,\,
\beta_n(k) = \mathop{\int}_0^c q_1(x)\Psi_n(x,k)\bar\Psi_0(x,k) dx,
$$
$$
v_n(k) = -i \mathop{\int}_0^c {\Psi'}_n(x,k)\bar\Psi_0(x,k) dx.
$$

The constructions described above can be summarized in the following

\underline{\bf Proposition 3}. {\it The links between $\lambda_n(k)$ and
$\mu_n(k)$ have the form }

\begin{equation} \label{21}
\lambda_n(k) N_0(k) - 2v_0(k)\mu_n(k) = \hat{T}_n
\{\lambda_i, \mu_i\, N_i(k), v_i(k)\, (i\leq
n-1),\beta_{n-1}(k)\}.
\end{equation}
{\it where $\hat{T}_n$ are some nonlinear functions of its
arguments which can be calculated explicitly for arbitrary $n$.
In particular, the first two functions have the form}
$$
\hat {T}_1\, =\, \beta_0(k),
$$
$$
\hat {T}_2\, =\, \beta_1(k) + \mu_1^2
N_0 - \lambda_1 N_1 +2\mu_1v_1.
$$

\underline{\it Remark}. To calculate $\hat{T}_n$ for arbitrary
$n$ one has to insert $R_n(x,k)$ which are known explicitly
into the solvability condition (\ref{18}).

Eqs.(\ref{14}) and (\ref{21}) give the $2\times 2$ linear system for the
functions $\lambda_n(k)$ and $\mu_n(k)$. By solving this
system one obtains the following formulae.

{\bf LEMMA 1}. {\it The correction terms of expansions (\ref{8}) and (\ref{9})
for the quasimomentum $p$ and dispersion function $E(p)$
have the form}
\begin{equation} \label{22}
\lambda_n(k) = \frac{\hat{T}_n\dot{\lambda}(k) +
2v_0(k)T_n}{\dot{\lambda}(k)N_0(k) - 2v_0(k)},
\end{equation}
\begin{equation} \label{23}
\mu_n(k) = \frac{\hat{T}_n +
N_0(k)T_n}{\dot{\lambda}(k)N_0(k) - 2v_0(k)}.
\end{equation}
{\it Here $T_n$ and $\hat{T}_n$ are described in
propositions 1 and 2 respectively.}

Finally one has to find the correction terms in the series
(\ref{7}) by solving  the boundary value problem (\ref{15}), (\ref{16}).
Its solution can be represented as the linear combination of
a general solution of (\ref{17}) and a particular solution of
(\ref {15}). Thus
\begin{equation} \label{19}
\Psi_n(x,k) = B_n(k)\Psi_0(x,k) -\bar\Psi_0(x,k)
\mathop{\int}_{x_n}^{x}\bar\Psi_0^{-2}
dx'\mathop{\int}_{{x'}_n}^{x'}R_n\bar\Psi_0
dx''
\end{equation}
where $B_n(k)$, $x_n$ and ${x'}_n$ are some constants w.r.t. $x$.
These constants can be defined unambiguously by means of
quasiperiodic boundary conditions (\ref{16}) and the normalization
condition (\ref{12}). It gives
$
x_n \,= \,0,
$
$
B_n(k) \,=\, 0
$
and the constant ${x'}_n$ is defined by the equation
$
\mathop{\int}_0^c
dx'\bar\Psi_0^{-2}\mathop{\int}_{{x'}_n}^{x'}R_n  \bar\Psi_0
dx'' = 0.
$

The expressions (\ref{22}), (\ref{23}) and (\ref{19}) determine the expansions
(\ref{7}), (\ref{8}), (\ref{9}) completely.
However the correction terms of these expansions have
singularities at some
values of unperturbed quasimomentum $k_s$. As it has been said
already we call
the corresponding points $\lambda_s(k)$ as the singular
points.

{\bf LEMMA 2}. {\it There are two sets of singular points. One
of them $\{E_l^0\}_{l\geq 1}$ coincides with the ends of
bands of the unperturbed operator $H_0$. The singular
points $\{\varepsilon_{lN}^0\}$ of another set are located
inside bands of $H_0$. If the ratio of the perturbed
potential period to the unperturbed potential period is
$m\equiv c/a$ then inside the $N$-th band there are $(m-1)$
singular points distributed in such a way that the
corresponding values of quasimomentum divide the Brilloin
zone by $m$ equal segments, so}

\begin{equation} \label{24}
\varepsilon_{lN}^0 = \lambda\left({\frac{\pi
lN}{ma}}\right),\,\,\,\,l=1,...,m-1.
\end{equation}

\underline{\it Proof}. Firstly let us show that at the ends
of bands of the unperturbed operator $H_0$ the function $v_0(k)$
vanishes:
$v_0(k_N)=0,\, (k_N=\pi N/a)$. Indeed at the ends of bands
the Bloch functions $\Psi_0(x,k)$ and $\bar\Psi_0(x,k)$
coincide: $\Psi_0 = \bar \Psi_0 $. Hence
$$
v_0 = -i\mathop{\int}_0^c{\Psi'}_0\bar\Psi_0 dx =
-i\mathop{\int}_0^c{\Psi'}_0 \Psi_0 dx =
-\frac i2\left[{e^{2i\pi lm}\Psi_0^2(0,k_l) -
\Psi_0^2(0,k_l)}\right] = 0.
$$
As it has been mentioned already at the ends of bands $\dot{\lambda}(k_l)=0$.
Hence the denominator of fractions (\ref{22}), (\ref{23})  vanishes at
the ends of bands and the perturbation series (\ref{7}), (\ref{8}) and
(\ref{9}) are not valid at these points.

The second set of singular points is generated by the
singularities of $T_n$ which can be shown to be the same
as of $T_2$ given by (20) and hence are defined by the
equation
$
F'(\lambda) \equiv \frac{d}{d\lambda} F(\lambda)= 0.
$
Since $F(\lambda)= \cos kc$
one has that$
F'(\lambda) = -\sin kc \,\frac{dk}{d\lambda c}.
$
and hence $F'(\lambda)=0$ if $\sin kc =0$ or $dk/d\lambda
=0$. In the first case $k=k_{lN} = \pi lN/{ma}$,
$l=1,...,m,\,N\in{\bf Z}$. However if $l=m$ the
quasimomentum $k_{mN}$ corresponds to the end of band of the
unperturbed operator $H_0$ and the zero of $\sin k_{mN} c$
is compensated by the zero of $\dot{\lambda}(k_{mN})$. So we have to
consider only $k_{lN}$ with $l=1,...,m-1$. These
points divide each Brilloin zone by $m$ equal
segments. The corresponding
energy points $\varepsilon_{lN}^0 =\lambda(k_{lN})$,
$l=1,...,m-1$, are located inside the $N$-th band.

The points where $dk/d\lambda=0$ lie inside gaps. However
the singularities of $T_n$
at these points are compensated by the factor
$\dot{\lambda}(k)$ entering the fractions
(\ref{22}) and (\ref{23}). Q.E.D.

To complete the study of expansions (\ref{7}), (\ref{8}) and (\ref{9})
outside the singular
points one needs to analyze the properties of correction
terms $\lambda_n(k)$ and $\mu_n(k)$ when $k$ is inside or
outside the bands of the unperturbed
operator $H_0$. Remind that to the bands there correspond real values
of quasimomentum $k$, while in gaps quasimomentum has nontrivial
imaginary part: $k=\pi N/a +i\gamma$ ($N$ is the gap number).

{\bf LEMMA 3}. {\it At real $k$ ($k\neq\pi N/a,\, N=1,2,...$) the
corrections terms $\lambda_n(k)$ and $\mu_n(k)$ are real.
If $k=\pi N/a + i\gamma$ the correction terms
$\lambda_n(k)$ of the dispersion function are real while
the correction  terms $\mu_n(k)$ of quasimomentum are
pure imaginary.}

\underline{\it Remark}. The statement of lemma means that
outside  the singular
points the spectral bands and gaps of the unperturbed operator $H_0$
are stable under the perturbation. It means that if some point
$\hat{\lambda}$ ($\hat{\lambda}\neq
\varepsilon_l^0,\,\lambda\neq E_{lN}^0$) belongs to the band
(gap) of $H_0$ the shifted point $\hat E = \hat\lambda +
O(\delta)$ should also belong to the band (gap) of the
perturbed operator $H$. Indeed, if $k$ is real then
$\mu_n(k)$ are also real and the perturbed
quasimomentum $p$ given by expansion (\ref{8}) is also real. If
$k=\pi N/a + i\gamma$ then $\mu_n(k)$ are pure imaginary
and $p=\pi N/a + i(\gamma +O(\delta))$. Hence the
corresponding spectral point $E(p)$ belongs to the gap of
$H$. It should be noted that the number of
this gap is equal to $mN$ since the width of the
Brilloin zone of the perturbed operator is $\pi/c \equiv
\pi/ma$.

\underline{\it Proof}. Here we prove the statement of lemma for $n=1$.
For $n\geq 2$ the statement can be proven analogously. Remind that
$$
\lambda_1(k) = \frac{\beta_0(k)\dot{\lambda}(k)}
{\dot{\lambda}(k)N_0(k) - 2v_0(k)},
\,\,\,\,\,\,\,
\mu_1(k) = \frac{\beta_0(k)}
{\dot{\lambda}(k)N_0(k) - 2v_0(k)}.
$$

 To prove lemma we show:
1) that at real $k$ the quantities
$\dot{\lambda(k)},\,\beta_0(k)$, $N_0(k)$ and $v_0(k)$ are real;
2) if $k = \pi N/a +i\gamma$ the functions
$N_0(k)$ and $\beta_0(k)$ are real while the functions
$\dot{\lambda}(k)$  and $v_0(k)$ are pure imaginary.

Since $\cos ka = F_0(\lambda)$ where $F_0(\lambda)$
is the Lyapounov functions of the unperturbed operator $H_0$,
one has
$$
\dot{\lambda}\equiv \frac{d}{d\lambda} \lambda(k) =
-\frac{a \sin ka}{{F'}_0(\lambda)},\,\,\,{F'}_0(\lambda) =
\frac{d}{d\lambda} F_0(\lambda).
$$

Since the function ${F'}_0(\lambda)$ is always real at $\lambda
\in {\bf R}$, the
function $\sin ka$ is real at real $k$ and $\sin (N\pi +i\gamma a)
= (-1)^N i \sinh \gamma a$ is pure imaginary at $k=\pi N/a +i\gamma$,
then $\dot{\lambda}(k)$ is real inside bands and is pure
imaginary inside gaps.

       The fact that functions $N_0(k)$ and $\beta_0(k)$
are always real directly follows from their definitions.
Inside gaps ($k = \pi N/a +i\gamma$) the Bloch functions
are real.  Therefore the
function
$$
v_0(k) = -i \mathop{\int}_0^c {\Psi'}_0(x,k)\bar\Psi_0(x,k)dx
$$
is pure imaginary. At real $k$
$$
\bar v_0(k) = i \mathop{\int}_0^c
\bar \Psi'_0(x,k)\Psi_0(x,k)dx =
i\bar {\Psi'}_0\Psi_0\vert_0^c - i \mathop{\int}_0^c
{\Psi'}_0\bar\Psi_0dx =
$$
$$
= - i \mathop{\int}_0^c{\Psi'}_0\bar\Psi_0 dx \equiv v_0(k),
$$
so the function $v_0(k)$ is real. Q.E.D.

\section{ Local perturbation series at singular points}

       In this Section we construct the local perturbation series
at singular points and by the study of their properties prove two
basic statements. To formulate the first statement we need the
following notations. Let $m=c/a$ be the ratio of the perturbed
potential period to the unperturbed potential one. We introduce the
points $k_{lN} =\pi lN/ma,\,l=1,...,m-1$, where $N$ is
arbitrary integer. According to lemma 2
the points $\varepsilon_{lN}^0 = \lambda(k_{lN})$ are
singular. Let us introduce the functions
\begin{equation} \label{25.1}
\varphi_0(x) = \frac 12\left[{\Psi_0(x,k_{lN}) +\bar\Psi_0(x,k_{lN})}\right]
\end{equation}
and
\begin{equation} \label{25.2}
\hat \varphi_0(x) = \frac {1}{2i}\left[{\Psi_0(x,k_{lN}) -
\bar\Psi_0(x,k_{lN})}\right] ,
\end{equation}
where $\Psi_0(x,k)$ is the Bloch function of the
unperturbed operator $H_0$ fixed by
the conditions (\ref{11}). Consider the quantities
$\varepsilon_{lN}$ and $p_{lN}$  given up
to the second order of $\delta$ as follows
\begin{equation} \label{26.1}
\varepsilon_{lN} = \varepsilon_{lN}^0 +
\delta\varepsilon_{lN}^1+ O(\delta^2)
\end{equation}
and
\begin{equation} \label{26.2}
p_{lN} = k_{lN} +
i\delta\beta_{lN}^1+ O(\delta^2)
\end{equation}
Here
\begin{equation} \label{27}
\varepsilon_{lN}^1 =
\frac{\mathop{\int}_0^cq_1(x)\varphi_0^2(x)dx}
{\mathop{\int}_0^c\varphi_0^2(x)dx}
\end{equation}
and
\begin{equation} \label{28}
\beta_{lN}^1 =
\frac{-\varepsilon_{lN}^1{\mathop{\int}_0^c\varphi_0(x)
\hat{\varphi_0} dx} + {\mathop{\int}_0^c q_1(x)\varphi_0(x)
\hat{\varphi_0} dx}}{2\mathop{\int}_0^c\varphi_0^2(x)dx}.
\end{equation}
Under these notations the following statement is valid.

{\bf THEOREM 1}.{\it At the singular points
$\varepsilon_{lN}^0,\,l=1,...,m-1$  the $N$-th spectral band of the
unperturbed operator $H_0$ is splitted under the
perturbation $\delta q_1(x)$ into
$m$ bands. The separating gaps are concentrated around the points
$\varepsilon_{lN}$ given by (\ref{26.1}). The quasimomenta
$p_{lN}$ corresponding
to these points are complex and are described by (\ref{26.2}). The width of the
$lN$-th gap in the leading order of $\delta$ is}
\begin{equation} \label{T}
\Delta_{lN} \sim 2\delta|\beta_{lN}^1| + O(\delta^2)
\end{equation}
{\it with $\beta_{lN}^1$ given by eq.(\ref{28})}.

\underline{\it Proof}. To proof the statement we construct the local
perturbation series at the singular point $\varepsilon_{lN}^0$.
Denote by $\varepsilon_{lN}$ the point obtained from
$\varepsilon_{lN}^0$ under the perturbation $\delta q_1(x)$. Let us
seek this point in the form
\begin{equation} \label{29.1}
\varepsilon_{lN} = \varepsilon_{lN}^0 + \mathop{\sum}_{n\geq
1}\delta^n \varepsilon_{lN}^n.
\end{equation}

The corresponding value of quasimomentum $p$ we denote by $p_{lN}$
and also seek in the form of series
\begin{equation} \label{29.2}
p_{lN} = k_{lN} + \mathop{\sum}_{n\geq
1}\delta^n \mu_{lN}^n,
\end{equation}
where $k_{lN} = \pi lN/ma$ and
$\varepsilon_{lN}^0 = \lambda(k_{lN})$.

According to eq. (\ref{4}) the function $\Phi(x,p)$ related with the Bloch
function $\Psi(x,p)$ by eq. (\ref{2}) at the point $p=p_{lN}$ has the
property
\begin{equation} \label{30}
\Phi(x+c, p_{lN}) = e^{ik_{lN} c}\Phi(x, p_{lN}) =
(-1)^{lN}\Phi(x, p_{lN}).
\end{equation}
Thus $\Phi(x) \equiv \Phi(x,p_{lN})$ is periodic or
antiperiodic function.  Let us seek it in the
form
\begin{equation} \label{29.3}
\Phi(x) = \varphi_0(x) + \mathop{\sum}_{n\geq
1}\delta^n \varphi_n(x),
\end{equation}
where the function $\varphi_0(x)$ is given by (\ref{25.1}).
Note that due to (\ref{11})
the leading term $\varphi_0(x)$ is periodic or antiperiodic, i.e.
$
\varphi_0(x+c) = (-1)^{lN} \varphi_0(x).
$
To provide the property (\ref{30}) the correction terms
$\varphi_n(x)$ are to be
sought in a class of functions with the same boundary condition:
$
\varphi_n(x+c) = (-1)^{lN} \varphi_n(x).
$

After inserting the Bloch function $\Psi(x,p_{lN})= \exp
\{i(p_{lN}-k_{lN})x\} \Phi(x)$ into equation (\ref{6})
with $E = \varepsilon_{lN}$ one obtains for $\Phi(x)$
the following equation
$$
\left[{-\partial_x^2 - 2i(p_{lN} - k_{lN})\partial_x +(p_{lN}
- k_{lN})^2 + q_0(x) + \delta q_1(x) -
\varepsilon_{lN}}\right] \Phi(x) = 0.
$$
Now we insert the expansions (\ref{29.1}), (\ref{29.2}),
(\ref{29.3}) into the last equation and
equate terms at equal powers of $\delta$. In the leading order of
$\delta$ one obtains identity
$$
\left[{-\partial_x^2 + q_0(x) - \varepsilon_{lN}}\right]
\varphi_0(x) = 0.
$$
In the first order one has the equation
$$
\left[{-\partial_x^2 + q_0(x) + - \varepsilon_{lN}}\right]
\varphi_1(x) = \left[{\varepsilon_{lN}^1 - q_1(x)
+2i\mu_{lN}^1 \partial_x}\right] \varphi_0(x) \equiv R_1(x).
$$
Following the discussions of Sec.2 one concludes that the
condition of solvability of this equation in a class of periodic
(antiperiodic) functions is the orthogonality of $R_1(x)$
to periodic (antiperiodic) solutions
 of the corresponding homogeneous equation.
Since there are two such solutions (say, for example, $\varphi_0(x)$ and
$\hat \varphi(x)$ given by (\ref{25.1}) and (\ref{25.2}) respectively) one
has two solvability conditions
$$
\mathop{\int}_0^c R_1(x)\varphi_0(x) dx =0, \,\,\,\,\,\,\,
\,\,\,\,\,\,\mathop{\int}_0^c R_1(x)\hat{\varphi_0}(x) dx =0,
$$
where the fact that $\varphi_0$ and $\hat{\varphi_0}$ are real
is taken into account.
Since $\mathop{\int}_0^c {\varphi'}_0(x)\varphi_0(x)
dx = 0$, the first condition allows to define $\varepsilon_{lN}^1$
and yields relation (\ref{27}). The second condition leads to the relation
$
\mu_{lN}^1 = i \beta_{lN}^1
$
with $\beta_{lN}^1$ defined by (\ref{28}).

      Now one knows the r.h.s. $R_1(x)$ explicitly and can easily find
the periodic (antiperiodic) solution $\varphi_1(x)$. Then the whole
procedure can be repeated. Namely, one can write down the equation in
the second order of $\delta$ (equation for $\varphi_2(x)$) and
find the correction terms $\varepsilon_{lN}^2$ and $p_{lN}^2$
from two  solvability conditions
$$
\mathop{\int}_0^c R_2(x)\varphi_0(x) dx =0,
\,\,\,\,\,\,\,\,\,\,\,\,\,\,\,\,
\mathop{\int}_0^c R_2(x)\hat{\varphi_0}(x) dx =0.
$$
The described procedure allows to define the correction terms in
expansions (\ref{29.1}),(\ref{29.2}), (\ref{29.3})  of arbitrary order. However this is out the
scope of the present paper.

      Since the width of the Brilloin zone for the perturbed operator is
$\pi/c = \pi/ma$ then according to the general theory of
periodic operators the
complex quasimomentum $p_{lN} = \pi lN/ma + i\delta
\beta_{lN}^1 +O(\delta^2)$ corresponds to the real energy
$\varepsilon_{lN} = E(p_{lN})$ which lies
in the $lN$-th gap of $H$.

      To calculate the width of gaps appeared under the
perturbation let us calculate the Lyapounov function ${\cal F}(E)$ of the
perturbed operator at the point $E =\varepsilon_{lN}$:
$$
{\cal F}(\varepsilon_{lN}) = \cos (k_{lN} + i\delta
\beta_{lN}^1 + O(\delta^2)) =
(-1)^{lN}\cosh(\delta |\beta_{lN}^1| +O(\delta^2)) =
$$
$$
= (-1)^{lN}\left({1+ \frac{\delta^2|\beta_{lN}^1|}{2} +
O(\delta^4)}\right).
$$

Now let us expand the Lyapounov function in the vicinity of
$\varepsilon_{lN}$ in the Taylor series
${\cal F}(E) = {\cal F}(\varepsilon_{lN}) + \frac{{\cal
F}''(\varepsilon_{lN})}{2} (E - \varepsilon_{lN})^2 +
O((E -  \varepsilon_{lN})^3)$
and set $E = E_{lN}^{\pm}$ where $E_{lN}^{\pm}$ are the ends
of the gap. Here we used
the fact that ${\cal F}'(\varepsilon_{lN}) = 0$. Since the
number of  the considered gap is $lN$ one has
$$
{\cal F}(E_{lN}^{\pm}) = (-1)^{lN},\,\,{\cal
F}''(\varepsilon_{lN}) = (-1)^{lN+1}|{\cal F}''(\varepsilon_{lN})|.
$$
and hence the width of the $lN$-th gap in the leading order of $\delta$ is
$$
\Delta_{lN} \equiv (E_{lN}^+ - \varepsilon_{lN}) - (E_{lN}^- -
\varepsilon_{lN}) \sim 2\delta |\beta_{lN}^1|.
$$
Q.E.D.

     Now it remains to consider the singular points which coincide
with ends of bands (gaps) of the unperturbed operator.
Let us denote by $E_{N+}^{(0)} = \lambda (\pi N/a + 0)$ and
$E_{N-}^{(0)} = \lambda (\pi N/a - 0)$ the right and
left ends of the $N$-th  gap in the spectrum of the
unperturbed operator $H_0$. The following statement is valid.

{\bf LEMMA 4}. {\it Under the perturbation $\delta q_1(x)$ the
ends of the $N$-th gap of
the unperturbed operator are shifted as follows}
\begin{equation} \label{32}
E_{lN\pm} = E_{N\pm}^{(0)} + \delta E_{N\pm}^{(1)} +
O(\delta^2)
\end{equation}
{\it where}
\begin{equation} \label{33}
E_{N\pm}^{(1)} = \frac{\mathop{\int}_0^c q_1(x)\Psi_0^2(x)
dx}{\mathop{\int}_0^c \Psi_0^2(x)
dx}
\end{equation}
{\it and $\Psi_0(x)\equiv \Psi_0(x, \pi N/a \pm 0)$ is the Bloch
function $\Psi_0(x,k)$ of $H_0$ taken at the ends of the $N$-th gap.}

\underline{\it Proof}. To obtain the expansion (\ref{32}) let us
insert the series
\begin{equation} \label{34}
E_{lN\pm} = E_{N\pm}^{(0)} + \mathop{\sum}\limits_{n\geq
1}\delta^n E_{N\pm}^{(n)},\,\,\,\,\,\,\,
\Psi(x) = \Psi_0(x) + \mathop{\sum}\limits_{n\geq
1}\delta^n \psi_n(x)
\end{equation}
into equation (\ref{6}) considered at $E = E_{lN\pm}$ where
$E_{lN\pm}$ are the boundary
points of the perturbed operator spectrum. Here $\Psi(x)$ is the Bloch
function of $H$ at the point $E_{lN\pm}$.

     In the leading order of $\delta$ one obtains the identity
$$
\left[{-\partial_x^2 + q_0(x) - E_{N\pm}^{(0)}}\right]
\psi_0(x) = 0.
$$
The first order term has the form
\begin{equation} \label{34'}
\left[{-\partial_x^2 + q_0(x) - E_{N\pm}^{(0)}}\right]
\psi_1(x) = E_{N\pm}^{(1)}\psi_0(x) - q_1(x) \psi_0(x)
\equiv R_1.
\end{equation}
Since the functions $\Psi(x)$ and $\psi_0(x)$ are periodic
(antiperiodic)  with the
period $c$, one has to find the correction terms $\Psi_n(x)$ in the same
class. The solvability condition of equation (\ref{34'}) in such class of
functions is the orthogonality of $R_2$ to periodic (antiperiodic)
solutions of homogeneous equation. Since at the boundary points of
spectrum $\Psi_0(x,k) = \bar\Psi_0(x,k)$, the function $\Psi_0(x)$ is the
single  periodic (antiperiodic) solution and the
solvability condition has the form
$
\mathop{\int}\limits_0^cR_2(x)\Psi_0(x) dx = 0.
$
It yields the first correction term $E_{N\pm}^{(1)}$ in
the form (\ref{33}).

     Obviously the above described procedure can be repeated again
and allows to define all terms $E_{n\pm}^{(l)}$ in
expansion (\ref{34}). Q.E.D.

It is convenient to combine the statements of theorem 1 and
lemma 4 in the following form.

{\bf THEOREM 2.} {\it Let $c/a=m \in {\bf N}$ be the ratio of the perturbed
operator $H$ period to the unperturbed operator $H_0$ one. Then the gaps
in the spectrum of $H$  can be divided into two sets: the "old"
gaps with numbers $mN$, $N = 1,2,...$ which are just shifted
gaps of the unperturbed operator $H_0$ with the shifts given by
eqs.(\ref{32}),(\ref{33}); and "new" gaps with numbers $lN, l=1,2,...,m-1,
(N=1,2,...)$ which arise under
perturbation. These new gaps are located inside the $N$-th band of the
unperturbed operator, are concentrated around the points
$\varepsilon_{lN}$ given by eq.(\ref{26.1}) and have the widths
described by eq.(\ref{T}).}

\end{document}